# Aging and memory properties of topologically frustrated magnets

V. Dupuis[1], E. Vincent[1], J. Hammann[1], J. E. Greedan[2] and A. S. Wills[3]

1. Service de Physique de l'Etat Condensé, CEA Saclay, 91191 Gif-sur-Yvette cedex, France
2. Brockhouse Institute for Materials Research, Mc Master University,
   1280 Main Street West, Hamilton, ON, Canada
3. Institut Laue Langevin, BP156, 38042, Grenoble cedex, France

**Abstract**

The model 2d kagomé system $(H_3O)Fe_3(SO_4)_2(OH)_6$ and the 3d pyrochlore $Y_2Mo_2O_7$ are two well characterized examples of low-disordered frustrated antiferromagnets which rather then condensing into spin liquid have been found to undergo a freezing transition with spin glass-like properties. We explore more deeply the comparison of their properties with those of spin glasses, by the study of characteristic rejuvenation and memory effects in the non-stationary susceptibility. While the pyrochlore shows clear evidence for these non-trivial effects, implying temperature selective aging, that is characteristic of a wide hierarchical distribution of equilibration processes, the kagomé system does not show clearly these effects. Rather, it seems to evolve towards the same final state independently of temperature.

**Introduction**

The highly degenerate ground state of geometrically frustrated antiferromagnets gives rise to a variety of exotic behaviors depending on whether and how the degeneracy is lifted. The observed properties range from quantum liquids to glasses and ordered states. The degree of degeneracy depends on the specific lattice on which the spins are located. Some of the largest degrees of degeneracy exist on networks of corner-sharing triangles (2d kagomé lattice) or tetrahedra (3d pyrochlore lattice). Consequently, these lattices have been the subjects of much research. Theoretical and numerical calculations on such lattices with classical Heisenberg spins have shown the persistence of paramagnetism down to 0 K [1-5], even with the presence of a small degree of disorder. Only when a certain threshold value is exceeded does the system enter a spin glass-like state [1,5]. Despite this, experiment has shown that essentially non-disordered compounds with these kagomé or pyrochlore lattices do exhibit a spin freezing transition at some low temperature $T_f$ (much lower than the Curie-Weiss temperature), below which they develop magnetic irreversibilities and long time magnetic relaxation. A question therefore arises over how these glassy states are stabilized and whether the observed behaviors are really similar to those of conventional spin glasses, where frustration always coexists with a large degree of disorder.

Two well known topologically frustrated antiferromagnets that exhibit a low T freezing transition are hydronium jarosite $(H_3O)Fe_3(SO_4)_2(OH)_6$ (kagomé) ($T_f$ = 17.8 K) and the pyrochlore $Y_2Mo_2O_7$ ($T_f$ = 20.5 K). Both display several characteristic static and dynamic spin glass properties [6,7,8,9,10,11] despite the persistence of fast magnetic fluctuations at low temperatures [8,12].

In order to go deeper into the comparison between topologically frustrated systems and canonical spin glasses, we explore in this paper the detailed behavior of the non-stationary response function of both materials as a function of temperature. In particular we checked for the non-trivial but universal behavior of *rejuvenation and memory* found in site-disordered spin glasses [13]. The *rejuvenation* effect denotes the fact that the evolution of the system at any temperature $T_1$ below $T_f$ has little or no effect on the magnetic response function at any (sufficiently) different temperature $T_2$. More specifically, if the system after an initial quench has been allowed to evolve (age) only at $T_1$, it will display the initial quenched response function at any other temperature away from $T_1$. This is in contrast to a simple evolution towards a temperature independent equilibrium state. The *memory* effect relates to the fact that whatever the thermal history at $T<T_1$, the response function recovers its "aged" value upon re-heating to $T_1$. These effects are well illustrated in Fig.1 which reports zero-field cooled (ZFC) dc susceptibility results obtained after different cooling procedures (experimental protocol of ref[14]) on the well characterized insulating thiospinel spin glass $CdCr_{1.7}In_{0.3}S_4$ ($T_f$=16.7 K) [15]. The dotted line in Fig.1 represents the magnetization in a small field (H=3 Oe) as a function of increasing temperatures, after a zero field slow cooling from above $T_f$ down to the lowest temperature (sweeping rates of 1.5 mK/s). The full line shows the magnetization under the same measuring conditions, after zero field cooling with the same cooling rate as before but with two long time stops of $10^4$ s at 14 K and $3.10^4$ s at 9.5K. Fig. 2 displays the relative difference between the two ZFC curves. Sharp dips are observed at both temperatures at which the system has been allowed to age for long waiting times- the system has kept the memory of its evolution at 9.5 K and 14 K during the cooling procedure. It is notable that the aging at these temperatures does not affect the response at other temperatures and that the effect is very selective and well marked.

The same experiments have been performed on both the kagomé and the pyrochlore compounds and the results are reported in Figs. 3 and 4.

## Hydronium jarosite $(H_3O)Fe_3(SO_4)_2(OH)_6$

The 2d kagomé system $(H_3O)Fe_3(SO_4)_2(OH)_6$ has large antiferromagnetic nearest neighbor interactions that are characterized by a Curie-Weiss temperature $\theta \approx -1200$ K. It exhibits a freezing temperature of $T_f=17.8$ K below which the field cooled and zero-field cooled dc susceptibilities strongly separate and long time magnetic relaxations appear. The large depression of the freezing temperature with respect to the Curie-Weiss temperature (small value of $T_f/|\theta| = 0.015$) is a signature of its very strong frustration. No long range ordering was detected by neutron diffraction [7]. A detailed investigation of the dynamic properties of the frozen state was done with the same sample used in previous studies of the crystal structure, susceptibility and µSR spectroscopy [6,8]. The results indicated some remarkable similarities with spin glasses [9]:
i) the frequency dependence of the ac susceptibility peak is inconsistent with an Arrhénius law, but can be fitted to a critical scaling law with an exponent similar to that of spin glasses. This indicates a finite temperature critical transition at $T_f=17.8$ K.
ii) the slow magnetic relaxation is strongly non-stationary (aging effects). In other words, the relaxation obtained after a quench from above $T_f$ and a small change of magnetic field depends strongly on the waiting time $t_w$ between the quench and the change of field. The longer $t_w$, the slower the relaxation of the magnetization. Such behavior demonstrates a stiffening of the system with time. Moreover, the scaling properties of this $t_w$ dependence have been found to be the same as those of spin glasses, with a scaling exponent $\mu=0.9$ [15]. Despite this similarity, temperature cycling experiments, performed using protocols developed in spin glass studies, showed marked quantitative differences with seemingly no rejuvenation nor temperature selective memory effects.

In the present investigation we made more accurate measurements using the alternative experimental protocol described above with an applied field of H=50 Oe and a sweeping rate of 1.5 mK/s. The results are reported in Fig.3. The full circles correspond to the difference between the reference ZFC magnetization and that obtained in case of a stop at 15 K for $10^4$ s during the cooling process. The open circles show the results for a much longer waiting time of $3.10^4$ s at 8 K. In the first experiment a large difference appears close to the freezing temperature, but no effect is detected at 15 K where the system was allowed to age for a long time. In the second experiment a similar difference is observed close to $T_f$, along with a small effect at around 8 K that is due to the long waiting time at that temperature. The main effect of the long waiting times applied during the cooling process is the marked maximum in the separation of the ZFC magnetization at $T_f$ and the continued separation over a large T-range below $T_f$. This shows that the evolution of the system is not very selective in temperature, *i.e.* the aged state of the system is not very temperature dependent, in strong contrast to the behavior of conventional spin glasses. In order to check this assertion, an experiment was made in which the sample was cooled down continuously using a slower cooling rate (0.33 mK/s) than for the reference (1.5 mK/s). The shape of the difference curve (the full line in Fig. 3) is very similar to that of the other curves, which proves that over the entire frozen range the system can be brought closer to its equilibrium state by aging at *any* given temperature.

## The $Y_2Mo_2O_7$ pyrochlore

The 3d pyrochlore compound $Y_2Mo_2O_7$ has also strong nearest neighbor antiferromagnetic interactions with a Curie-Weiss temperature $\theta = -200$ K. A frozen state with magnetic irreversibilities appears below $T_f=20.5$K. The ratio $T_f/|\theta| = 0.11$ is much larger than for the jarosite, suggesting the existence of a more efficient mechanism to overcome the frustration. The material is stoichiometric and crystallographically well ordered. The spin freezing transition has been carefully studied [11] and found to be a true thermodynamic transition with a power law divergence of the non-linear susceptibility. This feature is considered as an essential characteristic of the spin glass phase.

We have performed experiments to study the dynamic properties of the compound. We found slow non-stationary relaxation functions similar to those of spin glasses. In particular, there is a strong dependence of the magnetic response to a small excitation on the time (waiting time $t_w$) spent in the frozen state prior to the application of the excitation and that corresponding to the beginning of the measurement. Furthermore, the scaling of the $t_w$ dependence previously mentioned for the jarosite is also found- the scaling exponent has a very similar value ($\mu \approx 0.8$).

In Fig.4, we report the results of the measurements made with the same general procedure as above in a field of 50 Oe and a sweeping rate of 4.2 mK/s. Quantitatively, in addition to the reference curve, the two following cooling protocols have been used. The first one includes a single stop at 15 K for $10^4$ s, the resulting difference in the subsequent ZFC magnetization as compared to the reference curve is shown by the open circles in Fig.4. The second protocol includes two long waiting times of $10^4$ s at 18 K and 12 K. The results correspond to the full circles in Fig.4.

Clear memory effects appear in both experiments at the temperatures where the system was allowed to evolve for a long time. Away from these temperatures the difference between the curves goes to zero, which indicates that the long stays at the selected temperatures have not brought the system closer to equilibrium in the whole range below $T_f$. This behavior is very close to that shown in Fig.1, even though there are some quantitative differences: for similar characteristic temperatures and waiting times the observed dips are only of order 1.5 % whereas 4 % is observed in the spin glass $CdCr_{1.7}In_{0.3}S_4$. A more significant difference appears in the comparison of the widths of the dips: these are larger in the case of the pyrochlore, implying a lesser temperature selectivity in the memory effect.

**Conclusion**

The pyrochlore $Y_2Mo_2O_7$ has been shown to have most of the specific qualitative characteristics of a conventional 3d spin glass, apart from the already mentioned persistence of fast fluctuations at low temperatures [12]. The reason for the occurrence of the well defined glassy state in this very pure compound with a high degree of frustration is unclear. The effect of possible site disorder has been considered and shown to be inefficient in removing the high degeneracy of the ground state [5]. While it has been suggested recently that some bond disorder might be present in the material [16], such a property is likely to be sample dependent and no evidence to support its existence has been found in our sample. Even if it is present, theoretical work has concluded that bond disorder alone is insufficient to explain glassy behavior in a system with isotropic spins [17]. Rather, it is only when a reduction in the spin dimensionality is also present that a 3d glassy state is stabilized.[18]

The 2d jarosite $(H_3O)Fe_3(SO_4)_2(OH)_6$ contrasts with the $Y_2Mo_2O_7$ pyrochlore and site-disordered spin glasses as it does not display distinct rejuvenation and memory effects. Instead, any long stay below the freezing temperature has a cumulative and unselective effect on the temperature dependent response function that is displayed in most of the frozen temperature range. This suggests that its glassy magnetic phase evolves towards an almost temperature independent state. Its phase space therefore seems to be of a lesser complexity than that of the pyrochlore. At present the origin of the freezing in this topologically frustrated system is not at all clear. We are currently exploring a possible mechanism that involves in-plane anisotropy [19].

**Figure captions**

Fig.1 - Magnetization M over magnetic field H versus temperature T in a $CdCr_{1.7}In_{0.3}S_4$ spin glass sample ($T_f$=17.6K). The measurements were made using zero field cooling procedures (ZFC): the sample was first cooled down to the lowest temperature, a constant field H=3Oe was then applied and the dc magnetization was measured as a function of increasing temperatures. The dotted curve is obtained using a constant cooling rate of 1.5mK/s during the cooling process. The full curve corresponds to a cooling procedure with the same sweeping rate of 1.5mK/s but with two long time stops, one at 14K for $10^4$s, the other at 9.5K for $3.10^4$s (see inset).

Fig.2 - Relative difference ($M_{aging}$ - $M_{ref}$)/$M_{ref}$ between the aging and reference ZFC curves of fig.1 magnified by a factor 100.

Fig.3 - Magnified relative differences ($M_{aging}$ - $M_{ref}$)/$M_{ref}$ between aging and reference ZFC curves for the hydronium jarosite $(H_3O)Fe_3(SO_4)_2(OH)_6$. The measurements were made in a field H=50 0e. The ZFC reference is obtained using a temperature sweeping rate of 1.5mK/s. The full curve shows the difference between a ZFC measurement with a slow cooling rate of 0.33mK/s and the reference. The full circles and the open circles correspond to aging ZFC curves with the same sweeping rate as the reference, but with single stops respectively at 15K for $10^4$s and at 8K for $3.10^4$s,.

Fig.4 - Magnified relative differences ($M_{aging}$ - $M_{ref}$)/$M_{ref}$ between aging and reference ZFC curves for the $Y_2Mo_2O_7$ pyrochlore. The measurements were made in a field H=50 0e and a temperature sweeping rate of 4.2mK/s. The open circles correspond to an aging ZFC curve with a single stop at 15K for $10^4$s and the full circles to the case of two stops, one at 18K, the other at 12K, for a waiting time of $10^4$s.

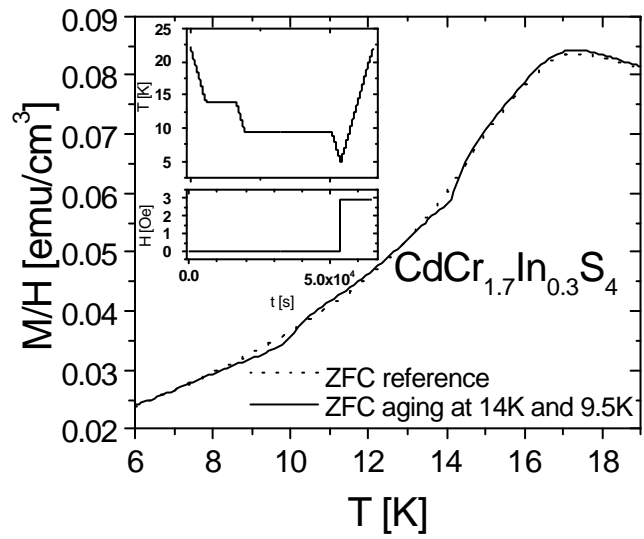

Fig. 1  V. Dupuis  J. A. P.

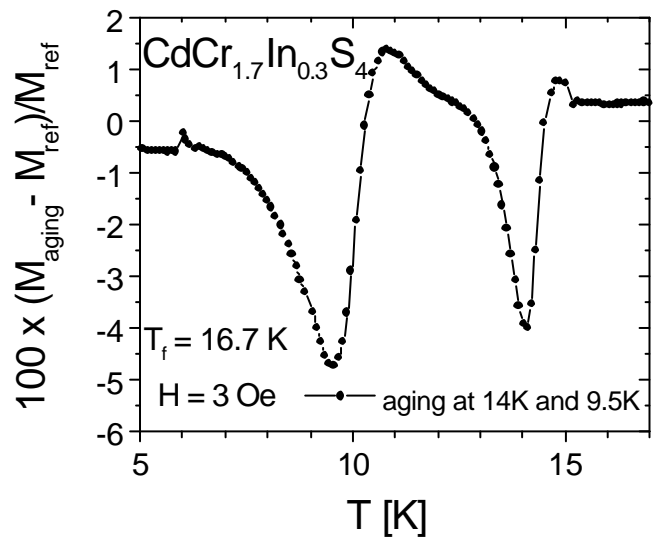

Fig. 2  V. Dupuis    J.A.P.

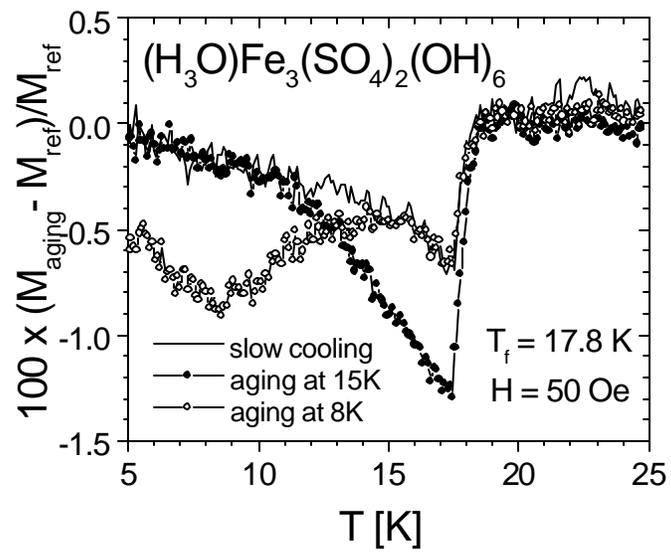

Fig. 3   V. Dupuis      J. A. P.

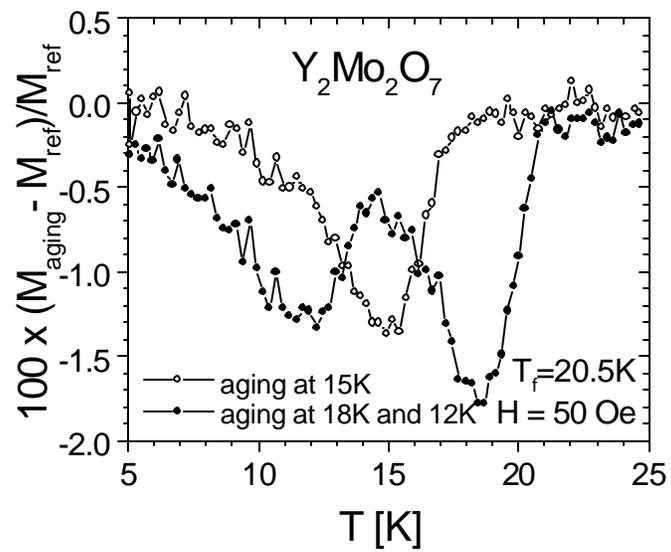

Fig. 4    V. Dupuis        J. A. P.